\documentclass{jps-cp}
\usepackage{txfonts} 
\usepackage{pdfpages}
\usepackage{graphicx}

\title{Nuclear Electric Dipole Moment as a Good Probe of CP Violation}

\author{Nodoka \textsc{Yamanaka}$^{1,2}$}

\inst{
$^{1}$IPNO, CNRS-IN2P3, Univ. Paris-Sud, Universit\'{e} Paris-Saclay, 91406 Orsay Cedex, France\\
$^{2}$Theoretical Research Division, Nishina Center, RIKEN, Saitama 351-0198, Japan
}

\email{
yamanaka@ipno.in2p3.fr
}

\recdate{February 1, 2019}

\abst{
The electric dipole moment (EDM) is an excellent probe of new physics beyond the standard model of particle physics. 
The EDM of light nuclei is particularly interesting due to the high sensitivity to the hadron level CP violation. 
In this proceedings contribution, we investigate the mechanism of the generation of the EDM for several light nuclei and the prospect for the discovery of new physics.
}

\kword{CP violation, electric dipole moment, light nuclei
}

\begin{document}
\maketitle

\def\vc#1{\mbox{\boldmath $#1$}}

\section{Introduction}

The CP violation of the standard model (SM) is insufficient to explain the observed matter/photon ratio $1:10^{10}$ \cite{sakharov,farrar,huet}.
The electric dipole moment (EDM) is an observable very sensitive to the CP violation, and its experimental measurement is expected to unveil the CP violation beyond the standard model (SM) \cite{pospelovreview,engeledmreview,yamanakabook,devriesedmreview,yamanakanuclearedmreview,atomicedmreview,chuppreview}.
It is becoming possible to measure the EDM of charged particles with very high precision using storage rings, and dedicated experiments are currently under development \cite{anastassopoulos,jedi,jedi2,bnl}.
In this proceedings contribution, we summarize the recent progress of the theoretical study of the EDM of light nuclei and the prospects on the discovery of new physics beyond the SM.

\section{The CP-odd Hamiltonian}

The nuclear EDM is induced by the CP-odd nuclear force.
We assume that the following one-pion exchange CP-odd nuclear force, with the isoscalar, isovector, and isotensor structures, is generated by the hadronic CP violation of the new physics \cite{yamanakanuclearedmreview,pvcpvhamiltonian2}:
\begin{eqnarray}
H_{P\hspace{-.35em}/\, T\hspace{-.5em}/\, }^\pi
& = &
-\frac{m_\pi}{8\pi m_N} 
\bigg\{ 
\bar{G}_{\pi}^{(0)}\,{\vc{\tau}}_{1}\cdot {\vc{\tau}}_{2}\, {\vc{\sigma}}_{-}
+\frac{1}{2} \bar{G}_{\pi}^{(1)}\,
( \tau_{+}^{z}\, {\vc{\sigma}}_{-} +\tau_{-}^{z}\,{\vc{\sigma}}_{+} )
+\bar{G}_{\pi}^{(2)}\, (3\tau_{1}^{z}\tau_{2}^{z}- {\vc{\tau}}_{1}\cdot {\vc{\tau}}_{2})\,{\vc{\sigma}}_{-} 
\bigg\}
\cdot
\frac{ \vc{r}}{r} \,
\nonumber\\
&& \times
\frac{e^{-m_\pi r }}{r} \left( 1+ \frac{1}{m_\pi r} \right)
.
\label{eq:CPVhamiltonian}
\end{eqnarray}
Here $\vc{r} \equiv \vc{r}_1 - \vc{r}_2$, ${\vc{\sigma}}_{\pm} \equiv {\vc{\sigma}}_1 \pm{\vc{\sigma}}_2$, and ${\vc{\tau}}_{\pm} \equiv {\vc{\tau}}_1 \pm{\vc{\tau}}_2$ denote the coordinate, spin, and isospin, respectively, with subscripts 1 and 2 labeling the nucleon.

\section{The Nuclear electric dipole moment}

The nuclear EDM induced by the CP-odd nuclear force is defined by
\begin{eqnarray}
d_{A}^{\rm (pol)} 
&=&
\sum_{i=1}^{A} \frac{e}{2} 
\langle \, \tilde \Phi_J (A) \, |\, (1+\tau_i^3 ) \, r_{iz} \, | \, \tilde \Phi_J (A) \, \rangle
,
\label{eq:polarizationedm}
\end{eqnarray}
where $|\, \tilde \Phi_{J} (A)\, \rangle$ denotes the polarized nuclear state.
Since the CP-odd nuclear couplings are small, the nuclear EDM can be parametrized as a linear combination of them:
\begin{eqnarray}
d_{A}^{\rm (pol)} 
&=&
\bar G_\pi^{(0)}
a_\pi^{(0)}
+\bar G_\pi^{(1)}
a_\pi^{(1)}
+\bar G_\pi^{(2)}
a_\pi^{(2)}
.
\label{eq:polarizationedm}
\end{eqnarray}
Evidently, nuclei with large coefficients $a_\pi^{(i)}$ $(i=0,1,2)$ are the most interesting.
In Table \ref{table:nuclearedm}, we summarize known results of nuclear EDM.

\begin{table}[tbh]
\caption{
Coefficients of Eq. (\ref{eq:polarizationedm}).
The case of the neutron is also displayed for comparison \cite{crewther}.
The other coefficients of this table do not include the effect of the nucleon EDM.
}
\label{table:nuclearedm}
\begin{center}
\begin{tabular}{lccc}
\hline
  &$a_\pi^{(0)}$ ($10^{-2} e$ fm) & $a_\pi^{(1)}$ ($10^{-2} e$ fm) & $a_\pi^{(2)}$ ($10^{-2} e$ fm) \\ 
\hline
$n$ \cite{crewther} & $1$ & $- $ & $-1$  \\
$^{2}$H \cite{liu,yamanakanuclearedm} & $-$ & $1.45 $ & $-$  \\
$^{3}$He \cite{bsaisou,yamanakanuclearedm}& $0.59$ & 1.08 & 1.68  \\
$^{3}$H \cite{yamanakanuclearedm} & $-0.59$ & 1.08 & $-1.70$  \\
$^{6}$Li \cite{yamanakanuclearedm}& $-$ & 2.2 & $-$  \\
$^{7}$Li \cite{Yamanaka:2018dwa} & $-0.6$ & 1.6 & $-1.7$  \\
$^{9}$Be \cite{yamanakanuclearedm}  & $-$ & $1.4$ & $-$  \\
$^{11}$B \cite{Yamanaka:2018dwa} & $-0.5$ & $2$ & $-1$  \\
$^{13}$C \cite{c13edm} & $-$ & $-0.20 $ & $-$  \\
$^{129}$Xe \cite{yoshinaga2} & $7.0\times 10^{-3}$ & $7.4\times 10^{-3}$ & $3.7\times 10^{-2}$  \\
\hline
\end{tabular}
\end{center}
\end{table}

The EDM of the deuteron, $^3$He, and $^3$H has been calculated ab initio with consistent results \cite{liu,yamanakanuclearedm,bsaisou}.
That of p-shell nuclei has been evaluated in the cluster model \cite{yamanakanuclearedm,c13edm,Yamanaka:2018dwa}.
The results for $^6$Li, $^7$Li, $^9$Be, and $^{11}$B are obeying an approximate counting rule 
\begin{equation}
d_A
= 
[{\rm cluster} \, (^2{\rm H} \, {\rm or}\, ^3{\rm H}) \, \mbox{EDM}]
+
N_{\alpha N} \times (0.005 - 0.007) \bar G_\pi^{(1)} e\, {\rm fm}
,
\end{equation}
where $N_{\alpha N}$ denotes the number of $\alpha - N$ subsystems with open nucleon spin.
We note that this formula does not apply for $^{13}$C which has bad overlaps between opposite parity states \cite{c13edm}.
The important point of the coefficients of Table \ref{table:nuclearedm} is the non-alignment which will allow us to constrain multiple parameters of the new physics.
We also see an enhancement of the sensitivity to $\bar G_\pi^{(1)}$ for $^6$Li and $^{11}$B, which is explained by the constructive interference of the CP-odd $\alpha -N$ interaction growing with the nucleon number.
This amplification is however limited by the destructive interference due to the configuration mixing relevant in large nuclei (see the result for $^{129}$Xe, Table \ref{table:nuclearedm}) \cite{yoshinaga2}.

\section{Prospects for new physics}

\begin{figure}[htb]
\begin{center}
\includegraphics[clip,width=0.6\columnwidth]{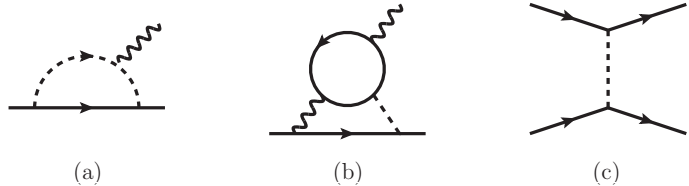}
\caption{
Elementary level CP violating processes, with (a) one-loop level fermion EDM diagram, (b) two-loop level Barr-Zee type diagram, and (c) tree level CP-odd four-quark interaction.
}
\label{fig:elementary_cpv}
\end{center}
\end{figure}

In many candidates of new physics beyond the SM, the EDM of light nuclei is induced at the tree, one-loop, or two-loop level (see Fig. \ref{fig:elementary_cpv}).
The representative model generating the one-loop level CP violating processes is the supersymmetric SM.
The quark EDM and the quark chromo-EDM [see Fig. \ref{fig:elementary_cpv} (a)] were extensively studied \cite{pospelovreview,ellismssmedm,ibrahimmssmedm,mssmreloaded}.
The quark EDM and chromo-EDM are suppressed by a factor of light quark mass, so the experimental constraint is looser than the naive dimensional analysis.
The prospective experimental sensitivity $O(10^{-29})e$ cm will probe the supersymmetry breaking scale of $O$(TeV).

In models such as the extended Higgs models \cite{barr-zee,Jung:2013hka,Brod:2018lbf}, supersymmetric models with large $\tan \beta$ \cite{pospelovreview,chang,Lebedev:2002ne}, R-parity violation \cite{rpv1,rpv2,rpv3,rpvlinearprogramming} or with heavy sfermions (split-supersymetry) \cite{splitsusy1}, the two-loop level Barr-Zee type diagram [see Fig. \ref{fig:elementary_cpv} (b)] yields the leading contribution to the nuclear EDM.

The four-quark interaction is generated at the tree level [see Fig. \ref{fig:elementary_cpv} (c)] in left-right symmetric model \cite{Xu:2009nt,deVries:2012ab,Maiezza:2014ala,Dekens:2014jka}.
From a simple estimation, the measurement of the nuclear EDM with the sensitivity of $O(10^{-29})e$ cm can probe the mass scale of the right-handed $W$ boson up to PeV.

The SM model contribution to the quark EDM and chromo-EDM is induced at the three-loop level with a value of the order of $d_q \sim 10^{-35} e$ cm \cite{czarnecki}, being thus negligible.
We have to note, however, that the long distance contribution at the hadronic level is three to four orders of magnitude larger \cite{smneutronedmmckellar,seng,yamanakasmcpvnn,yamanakasmdeuteronedm,Lee:2018flm}, which is not ``extremely'' small, although it is still below the prospective expermental sensitivity.

\section{Summary}

In this proceedings contribution, we studied the EDM of several nuclei and the prospects for the discovery of new physics beyond the SM.
The nuclear EDM, depending on the nuclear structure, may enhance or suppress the hadron level CP violation.
Light nuclei are in the best position to probe the strong CP violation, since heavy nuclei are affected by the configuration mixing, and also because of the experimental difficulty to handle heavy ions in storage rings.
The expected experimental sensitivity of $O(10^{-29})e$ cm may unveil many attractive candidates of new physics at the TeV-PeV scale.
With these arguments, we definitely recommend the realization of new experiments aiming for the measurement of the EDM of light nuclei.

\end{document}